\documentclass[12pt]{article}
\usepackage{amssymb}
\usepackage{graphicx}
\usepackage{amsmath}
\newcommand{\be}{\begin{equation}}
\newcommand{\ee}{\end{equation}}
\begin{document}
\baselineskip18pt
\title{Modular Schr\"{o}dinger equation and dynamical  duality}
\author{Piotr Garbaczewski\thanks{
Electronic address: pgar@uni.opole.pl}\\
Institute of Physics,  University  of Opole, 45-052 Opole, Poland}
\maketitle
\begin{abstract}
We discuss quite surprising properties  of the one-parameter  family  of modular
(Auberson and Sabatier (1994)) nonlinear
  Schr\"{o}dinger equations. We  develop a  unified  theoretical framework for this  family.
  Special attention is paid  to  the   emergent \it  dual \rm  time evolution scenarios which,
  albeit   running in the \it real time \rm parameter of the pertinent nonlinear equation,  in each considered case,
  may be mapped among  each other by means of a suitable   analytic continuation in time procedure.
    This dynamical duality is  characteristic for   non-dissipative
  quantum motions and their dissipative (diffusion-type processes) partners, and naturally  extends to  classical motions
   in confining and
  scattering potentials.
 \end{abstract}\noindent
\vskip0.2cm

\section{Motivation}

An inspiration for the present paper comes from the recent publication \cite{brenig} discussing  effects of
 various scale transformations upon the free Schr\"{o}dinger picture dynamics. In particular, it has been noticed that
 an appropriate definition of the scale covariance induces Hamiltonians which mix,  with a pertinent  scale exponent as
 a hyperbolic rotation angle, an original free quantum dynamics with its  free dissipative counterpart
  (effectively, a suitable version of the free Brownian motion), \cite{brenig} c.f. also  \cite{gar2}.

 The  two disparately different time evolution patterns  do  run with respect to the same real time  label. However, the ultimate
   "mixing" effect of the above mentioned scale transformations
  takes the form of the  the Lorentz-like
  transformation  Eq. (35)  where one Hamiltonian   takes  the role of a regular "time"  while another of the "space" dimension
  labels.
  This  obvious   affinity with the   Euclidean "space-time" notion and the involved  complex analysis methods
  (e. g. Wick rotation, imaginary time transformation, analytic continuation in time) sets both  conceptual  and possibly phenomenological
  obstacles/prospects pertaining to an existence of such dynamical patterns in nature.

  Most puzzling for us is the apparently dual   notion of the time label which
  while  a priori  referring to the real time evolution, may as well be interpreted as a Euclidean (imaginary time) evolution.
 It is our aim to address an issue in its full generality, by resorting to   a one-parameter family of modular  Schr\"{o}dinger equations,
 where  external conservative potentials admitted.

  The adopted perspective relies on standard  approaches to nonlinear dynamical systems where
 a sensitive  dependence on a control (here, coupling strength) parameter may arise, possibly   inducing  global changes of properties  of
 solutions to the equations of motion. That is exactly the case in the present analysis.
   We wish to demonstrate that a duality property (realized by  an imaginary time transformation) does relate solutions
 of  the  modular  nonlinear Schr\"{o}dinger equation   for various coupling constant regimes.

   In Section 2  we set a general Lagrangian and Hamiltonian framework for
  the subsequent discussion and indicate   that effectively the dynamics can be reduced to three specific coupling value choices, \cite{sabatier,goldin},
   each of them being \it separately \rm  discussed in the literature.
   We aim at a stationary   action principle formulation of  the modular Schr\"{o}dinger dynamics  and the related
   hydrodynamics (the special cases  are: the familiar Bohm-type quantum hydrodynamics \cite{holland} and the hydrodynamical picture of
   diffusion-type processes \cite{gar2}).
  Basic principles of the action principle workings are patterned after the classical  hydrodynamics treatises, \cite{fetter,goldstein}.

In Section 3,  in  Hamiltonian  dynamics  terms,  we give  a new derivation and further generalize the original arguments  of \cite{brenig}.
We employ scaling properties of the Shannon entropy of a continuous probability density $\rho \doteq \psi ^* \psi $. The scale covariant patterns
of evolution, non-dissipative  with an  admixture  of a dissipative component, are  established  in external  potential fields.

Section 4 is devoted to a detailed analysis  of the  emergent  \it time duality \rm notion   for the three special values $0,1,2$ of the coupling
parameter. There,  we pay more attention to a specific intertwine between  confining and scattering dynamical systems. A properly addressed
\it sign \rm issue  for the potential function appears to be vital for a mathematical consistency of the formalism (links with the theory of
dynamical semigroups that underlies the dissipative dynamics scenarios).   As a byproduct of a discussion we establish the Lyapunov functionals
 for  the considered dynamical patterns of behavior  and the  (Shannnon) entropy production time rate is singled out.

 Section 5 addresses more specific problems
that allow to grasp the duality concept from a  perspective of the theory of  classical conservative  systems and  diffusion-type
(specifically - Smoluchowski) stochastic  processes.
 A number of illustrative examples is worked out in detail, with an
  emphasis  on the time duality notion in the classically inspired (standard Hamilton-Jacobi equations) and dissipative patterns of evolution
  (modified Hamilton-Jacobi: e.g. Smoluchowski diffusion processes vs quantum dynamics).

\section{Modular Schr\"{o}dinger equations}

Let us consider a   subclass of  so-called  modular Schr\"{o}dinger equations
\cite{sabatier,goldin} which are  local and homogeneous nonlinear
generalizations of the  standard  Schr\"{o}dinger equation:
\begin{equation}
i\hbar \partial _t \psi  = \left[ - {\frac{\hbar ^2}{2m}} \Delta +
V \right] \psi   +   \left[ \kappa\, {\frac{\hbar ^2}{2m}}\,
 {\frac{ \Delta |\psi |}{|\psi |}}\right]\, \psi   \, ,
\, \label{one}
\end{equation}
where $\psi (x,t)$ is a  complex function,    $|\psi |
  \doteq (\psi ^* \psi )^{1/2}$,      $V(x)$  is a real function
   and a coupling parameter $\kappa$ is non-negative, $\kappa\geq 0$ .

If  $\kappa>0$,   the pertinent  nonlinear dynamics  is
known to preserve the $ L^2(R^n)$ norm of any initially  given
$\psi $, but not the Hilbert space  scalar product $(\psi , \phi
)$  of two different, initially given $\psi $ and $\phi $. The
induced dynamics is non-unitary in $L^2(R^n)$; unitarity is restored if $\kappa=0$. The pertinent  set of solutions is
a subset of $L^2(R^n)$, but not a linear subspace (no superposition principle for $\kappa \neq 0$).

For all $\kappa\geq 0$ a usual continuity equation holds true with the
familiar quantum mechanical definition of a (probability) density
current: $\partial _t \rho = - \nabla \cdot j$, where $\rho = \psi
^* \psi $ and $j= (\hbar /2mi)(\psi ^* \nabla \psi - \psi \nabla \psi ^*)$.
We   consider  normalized solutions only,
which sets a standard  form of    $j\doteq \rho \cdot v$ ,
 where $v= (\hbar /2mi)[ (\nabla \psi /\psi ) -  (\nabla \psi ^*/\psi ^*)] \doteq (1/m) \nabla s
 $ is regarded as a gradient velocity field and $\rho (x,t)= |\psi |^2(x,t)$  is a probability density
 on $R^n$.

\subsection{Lagrangian formalism}

The  one-parameter family of   modular  equations (\ref{one}), together  with its complex
conjugate version,  derives from the local Lagrangian  density ${\cal{L}}$ by
means of the stationary action principle \cite{takabayasi}-\cite{holland}.
  Let us consider   a functional of
  $\psi $-functions, their space and time derivatives, including complex conjugates:
$I[\psi , \psi ^*]  = \int_{t_1}^{t_2} L(t) dt$ where $L(t) = \int
{\cal{L}}(x,t) \, dx$, (we leave unspecified, possibly infinite,
integration volume).

We  impose  the stationary action condition  $\delta I[\psi ,\psi
^*] =0$   for independent variations $\delta \psi $, $\delta \psi ^*$
which are bound to vanish at integration volume boundaries.   By invoking
 elements of the  pedestrian   functional calculus \cite{holland}, like   e. g.
$\delta {\cal{L}}/\delta \psi  \equiv  \partial {\cal{L}}/\partial
\psi - \sum _i \nabla _i [\partial {\cal{L}}/\partial (\nabla _i\psi
)]$, one  ends up with the Euler-Lagrange equations:  $\partial
_t[\partial {\cal{L}}/\partial (\partial _t\psi ^*)]= \delta
{\cal{L}}/\delta \psi ^*$    and $\partial _t[\partial
{\cal{L}}/\partial (\partial _t\psi )]= \delta {\cal{L}}/\delta \psi
$.

If we properly  specify  the Lagrangian density ${\cal{L}}\doteq  {\cal{L}}_{\kappa}$:
  \begin{equation}
  {\cal{L}}_{\kappa}(x,t)  =  {\frac{i\hbar }{2}} [\psi ^* (\partial _t\psi ) - \psi (\partial _t\psi ^*)]    - {\frac{\hbar ^2}{2m}}
  \nabla \psi  \cdot \nabla \psi ^*  -  V (x)\,  \psi \, \psi ^*  +
  \label{lagrangian}
\end{equation}
$$
    \kappa\, {\frac{\hbar ^2}{8m}} \left[  {\frac{ \nabla \psi ^*}{\psi ^*}}   +
     {\frac{\nabla \psi }{ \psi }} \right]^2  \,  \psi \, \psi ^*   \, .
$$
the stationary action principle  yields a   pair of adjoint
modular equations which comprise  Eq.~(\ref{one}) in conjunction with
its complex conjugate:
\begin{equation}
-  i\hbar \partial _t \psi ^*  = \left[ - {\frac{\hbar ^2}{2m}}
\Delta     +   V \right] \psi ^*  +   \left[ \kappa\,
   {\frac{\hbar ^2}{2m}}\,   {\frac{ \Delta |\psi |}{|\psi |}}\right]\, \psi ^* \, .
\,   \label{two}
\end{equation}

We  have previously introduced a  current  velocity field $v(x,t)$ by means of   $j\doteq \rho \cdot v$. Our gradient
assumption $v=(1/m)\nabla  s$ follows from an implicit reference to
 the familiar Madelung  substitution:  $\psi =|\psi | \,
 \exp(is/\hbar)$, where $|\psi |^2=\rho $.  Its consequent  exploitation allows us to rewrite
 the Lagrangian density  (\ref{lagrangian}) as follows (we leave intact an original  order of respective  entries):
 \begin{equation}
  {\cal{L}}_{\kappa }(x,t)  =  - \rho \left[ \, \partial _ts
      + {\frac{m}{2}} (u^2 + v^2)\,    +  V (x)\,     -
  \kappa\, {\frac{m}2}\, u^2 \, \right]  \label{lagr}
 \end{equation}
 where $u(x,t) \doteq (\hbar / 2m)\, \nabla \rho /\rho  $ is another velocity field (named an osmotic velocity).

 Here,  $\delta I[\rho , s]=0$ gives rise to  a continuity equation $\partial _t\rho = - \nabla (\rho \, \cdot  v)$  and  yields:
\begin{equation}
\partial _ts  +   {\frac{1}{2m}} (\nabla s)^2 + V + (1-\kappa)\, Q   = 0 \, ,  \label{HJ}
\end{equation}
where, in view of   $|\psi |=  \rho ^{1/2}$,   the  familiar notion of the
de Broglie-Bohm quantum potential,  \cite{holland,wilhelm}, naturally appears:
\begin{equation}
Q  = Q(x,t)\doteq
  - {\frac{\hbar ^2}{2m}}\, {\frac{\Delta \rho ^{1/2}}{\rho ^{1/2}}} = - {\frac{\hbar ^2}{4m}}\,
   \left[ {\frac{\Delta  \rho }{\rho }} - {\frac{1}2}\left( {\frac{\nabla \rho }{\rho }}\right)^2 \right]                   \, .
\end{equation}
The modular  Schr\"{o}dinger equation (1) takes the form:
\begin {equation}
 i\hbar \partial _t\psi = [-(\hbar ^2/2m) \Delta + V]\psi  - \kappa Q\, \psi \, .
\end{equation}

 \subsection{Hamiltonian formalism}

 A symplectic structure can be associated with the  dynamical   system  (1)-(3) by  introducing  fields
 $\pi _{\psi }$ and  $\pi _{\psi ^*}$ that are conjugate to $\psi $ and $\psi ^*$ respectively:
  $\pi _{\psi }= \partial {\cal{L}}/\partial (\partial _t\psi )= (i\hbar /2)\psi ^*$ and
$\pi _{\psi ^*}= \partial {\cal{L}}/\partial (\partial _t\psi ^*)= -(i\hbar /2)\psi $.   The  subsequent  Legendre-type transformation
 defines  the Hamiltonian density:
 \begin{equation}
{\cal{H}}_{\kappa } =   \pi _{\psi } \cdot  \partial _t \psi   +  \pi _{\psi ^*} \cdot \partial _t \psi ^* - {\cal{L}}_{\kappa }
  ={\frac{\hbar ^2}{2m}}
  \nabla \psi  \cdot \nabla \psi ^* + \left[ V   - \kappa\, {\frac{\hbar ^2}{8m}} \left(  {\frac{ \nabla  \psi ^*}{\psi ^*}}   +
     {\frac{\nabla \psi }{ \psi }} \right)^2 \right]  \psi \, \psi ^* =  \,  \label{hamiltonian}
 \end{equation}
$$
\rho \left[{\frac{m}{2}}\,  v^2\,    +  V    +(1 -
  \kappa )\, {\frac{m}2}\, u^2\right]   =  \pi _{s} \partial _ts   - {\cal{L}}_{\kappa }
$$
where, this time with respect to the  polar  fields  $\rho (x,t)$ and $s(x,t)$,  we have:
$\pi _{\rho } = \partial {\cal{L}}/\partial (\partial _t\rho ) = 0$ and
$\pi _{s} = \partial {\cal{L}}/\partial (\partial _ts ) = -  \rho $.

    The  Hamiltonian reads
$H_{\kappa }(t)= \int  {\cal{H}}_{\kappa }(x,t)\, dx$.
For the variational calculus it is not ${\cal{L}}$ but $L=\int   {\cal{L}}\, dx$ that really matters. Therefore,
 it is useful to note that
\begin{equation}
L(t) = - \langle \partial _ts\rangle - H_{\kappa }(t) \, ,
\end{equation}
where,  in view of  $\int \rho \, dx = 1$, we can  introduce the mean value  $\langle \partial _ts\rangle  = \int  \rho \,
\partial _ts\, dx$.

Let us evaluate  the mean value of the generalized Hamilton-Jacobi equation (\ref{HJ}).
 By assuming a proper behavior  of $\rho $ at the  integration volume boundaries  \cite{wilhelm},
 we  readily get $\langle Q\rangle  \doteq \int Q\,  \rho \, dx= +(m/2)\langle u^2 \rangle > 0$.
Thence,  on dynamically admitted fields  $\rho (t)$ and $s(x,t)$,   $L(t) \equiv 0$,  i. e.
$\langle \partial _t s\rangle  =- H_{\kappa } $.

Let us consider two function(al)s $A=\int {\cal{A}}(x,t)\, dx$ and $B=\int {\cal{B}}(x,t)\, dx$, which may explicitly
 depend on time $t$.
We  define their Poisson bracket
\begin{equation}
\{ A,\, B \} = - {\frac{i}{\hbar}} \int dx\, \left( {\frac{\delta A}{\delta \psi}}  \,  {\frac{\delta B}{\delta \psi ^*}}
-    {\frac{\delta A}{\delta \psi ^*}} \,    {\frac{\delta B}{\delta \psi}}  \,  .
\right)
\end{equation}

In particular, while identifying $A\equiv \psi (x,t)$ and $B\equiv H_{\kappa }(t)$, we get the modular
 Schr\"{o}dinger equation (\ref{one}) in the form
\begin{equation}
\partial _t\psi = \{\psi , \, H_{\kappa } \}
\end{equation}
while, by setting $A\equiv \psi ^*$,  an adjoint equation arises
\begin{equation}
\partial _t\psi ^* =    \{\psi ^*, \, H_{\kappa } \}    \,  .
\end{equation}
We recall e. g. that $\dot{\pi }_{\psi }=  - \delta H_{\kappa }/\delta \psi $  while
$\dot{\psi } =\delta H_{\kappa }/\delta \psi $.

Since the time dependence of $H_{\kappa }(t)$ is realized only through the canonical fields,  the Hamiltonian surely
 is a constant of motion. Thence  $\langle \partial _t s\rangle  $ as well.

The polar decomposition $\psi = \rho ^{1/2} \exp (is/\hbar )$,  $\psi ^*= \rho ^{1/2} \exp (-is/\hbar )$  preserves a
symplectic structure. In the self-explanatory notation there holds
\begin{equation}
\{ A,B\} \doteq \{ A,B\} _{\psi ,\psi ^*}  =   \{ A,B\} _{\rho  ,s}  \, \label{bracket}
\end{equation}
and thence:
\begin{equation}
\partial _t\rho = \{\rho , H_{\kappa }\}   = {\frac{\delta H_{\kappa }}{\delta s}} =
- {\frac{1}m} \nabla  \left( \rho \, \nabla s \right)
\end{equation}
while
\begin{equation}
\partial _ts = \{ s, H_{\kappa }\} = -  {\frac{\delta H_{\kappa }}{\delta \rho }} = - {\frac{1}{2m}} (\nabla s)^2 - V - (1-\kappa)\, Q \, .
\end{equation}
The result is valid for all $\kappa\geq 0$.

 Let  $G (t) = \int dx\, {\cal{G}}(x,t)$.  Clearly, c.f. \cite{goldstein}, if the time-dependence of $G$ is  realized
 only through canonically conjugate fields $\rho (x,t)$ and $s(x,t)$ (and their derivatives), then
 \begin{equation}
 {\frac{d\, G}{dt}} = \{ G,H_{\kappa} \}    \, . \label{recipe}
  \end{equation}
One should realize that a particular  time-dependence  pattern  of $G(t)$  critically  relies on  the chosen  parameter range for
 $\kappa \in R^+$.

\subsection{Reduction to  effective  $\kappa=0$, $1$ and $2$ self-coupling regimes}

In the  notation of Eq.~(\ref{lagr}), a  distinguished role  of   $\kappa=0$, $1$ or   $2$  coupling parameter values
 is   particularly conspicuous.
 The relevance of only these  three  $\kappa=0$, $1$,  $2$  parameter values   may be further enhanced.

  Namely,  let us  recall \cite{sabatier}  important
differences between  properties of solutions of Eqs.~(\ref{one}) and
(\ref{two}), depending on whether  $0< \kappa <1$, $\kappa=1$  or
$\kappa>1$.

{\bf (i)}  In the parameter range    $0< \kappa <1$  (in fact $0 \leq  \kappa <1$),
 if $\psi (x,t) = |\psi | \, \exp (is/\hbar )$
 actually  is a solution of (1), then $\psi '(x',t') = |\psi '|\,
\exp (is'/\hbar )$,  with $x'=x$, $t'= (1- \kappa)^{1/2} t$, $|\psi
'|(x',t') = |\psi |(x,(1-\kappa)^{-1/2}t')$  and $s'(x',t') = (1-
\kappa)^{-1/2}s(x,t)$, automatically solves  the linear
Schr\"{o}dinger equation:
\begin{equation}
i\hbar \partial _{t'} \psi ' = \left[ - {\frac{\hbar ^2}{2m}} \Delta
+   {\frac{1}{1- \kappa}}\,  V \right] \psi '  \, .
\end{equation}
The  scaling transformation replaces a nonlinear problem
by the linear one, albeit with a re-scaled potential.

{\bf (ii)} For a specific value $\kappa =1$ we encounter the formalism that  derives  (though actually  generalizing)
from the wave picture of  classical Newtonian mechanics, \cite{holland}.

{\bf (iii)}   In case of $\kappa>1$, a repetition of previous scaling steps,
provided we  replace  $(1-\kappa)^{1/2}$  by $(\kappa- 1)^{1/2}$
 in the pertinent formulas, results in the following outcome:
  $\psi '(x',t') = |\psi '| \, \exp (is'/\hbar )$  is  a solution of
  the nonlinear Schr\"{o}dinger equation
\begin{equation}
i\hbar \partial _t' \psi '  = \left[ - {\frac{\hbar ^2}{2m}} \Delta
+ {\frac{V}{\kappa-1}} \right] \psi '   +  2\,  \left[  \,
{\frac{\hbar ^2}{2m}}\,
 {\frac{ \Delta |\psi '|}{|\psi '|}}\right]\, \psi '
\, \label{three}
\end{equation}
Effectively, in the whole coupling parameter range $\kappa\in R^+$,
only the cases of $\kappa=0$, $\kappa=1$ and $\kappa=2$ form a   mutually exclusive
family, both on mathematical and physical grounds.
Since for $\kappa=0$ and $\kappa=2$, the above scaling transformations trivialize, while
 being irrelevant for  the  distinctively  "borderline"  case  of  $\kappa=1$,  we
can safely restore the notation of Eqs.~(\ref{one}) and
(\ref{two}).

In connection with the choice of $\kappa=2$, one more observation is  of utmost  importance,
 \cite{sabatier}, see also \cite{zambrini}-\cite{gar1}.
 Namely,  if a complex
function
\begin{equation}
\psi (x,t) = |\psi | \exp (is/\hbar)
\end{equation}
is a solution
of (\ref{one}) with $\kappa=2$, then  the {\it  real} function
\begin{equation}
\theta _*
(x,t)=|\psi |\, \exp (- s/\hbar )
\end{equation}
 is a solution of  the
generalized   (forward) heat equation
\begin{equation}
\hbar \partial _{t} \theta _*  = \left[ {\frac{\hbar ^2}{2m}}
\Delta +    V \right] \theta _*  \,
\label{four}
\end{equation}
with a diffusion coefficient $D=\hbar /2m$  and an external potential ${\cal{V}}=V/2mD$.  By setting $V=0$
we would arrive at the  standard heat equation $\partial _{t} \theta _*  = D\,\Delta \,  \theta _* $.

Another real function  $\theta  (x,t)=|\psi |\, \exp (+ s/\hbar )$ is a
solution of the  time-adjoint (backwards) version  of Eq.~(\ref{four}):
\begin{equation}
- \hbar \partial _{t} \theta   = \left[ {\frac{\hbar ^2}{2m}}
\Delta +    V \right] \theta  \, .~ \label{five}
\end{equation}

Note that  if we would have started from Eqs.~(\ref{four})  and (\ref{five})  with
 the purpose to arrive at the modular equation
(\ref{one})  with $\kappa=2$,  the  ill-posed   Cauchy problem would possibly become a serious obstacle.
That in view  of the  occurrence of the backwards parabolic equation.

This ill-posedness   might  be healed by invoking the theory of strongly continuous dynamical  semigroups,
\cite{zambrini,zambrini1,olk}.
To this end  we  need to  choose $V(x)$ to be a continuous
function that is bounded from above, so that  $V'=  -V$ becomes bounded
from below.
Then  Eq.~(\ref{four}) would  acquire a "canonical"  form of the
forward diffusion-type equation related to the contractive  semigroup operator  $\exp (- \hat{H}t/\hbar )$:
\begin{equation}
\hbar \partial _{t} \theta _*  =  - \hat{H}  \theta _*  =  \left[
{\frac{\hbar ^2}{2m}} \Delta   -   V' \right] \theta _*  \, .
\label{six}
\end{equation}
Eq.~(\ref{six}), together with its time adjoint
  \begin{equation}
  \hbar \partial _{t} \theta  = \hat{H}  \theta =   \left[-  {\frac{\hbar ^2}{2m}}
  \Delta   +    V' \right] \theta   \,
  \end{equation}
  stand for   principal dynamical equations of  so-called Euclidean quantum mechanics  \cite{zambrini1},
  while  falling  into a broader framework of  the Schr\"{o}dinger
boundary data and stochastic interpolation problem, \cite{zambrini,zambrini1} see e.g. also \cite{olk}.

Note, that if one  assumes
that the external potential $V'(x)$  is a continuous function that is bounded from below, then
 the  operator  $\hat{H}= - (\hbar ^2/2m)\Delta + V'$   is  (essentially) self-adjoint  in $L^2(R^n)$.
  So, we  have consistently   defined unitary transformations $\exp(-i\hat{H}t/\hbar )$  in $L^2(R^n)$ and
  $i\hbar \partial _t \psi = \hat{H}\psi$,  together with  plus $-i\hbar \partial _t \psi^* = \hat{H} \psi ^*$,
   ($\kappa=0$ case),
   as their local manifestations.

  A careful analysis reveals  \cite{zambrini,zambrini1}  that the standard Schr\"{o}dinger picture
    dynamics can be mapped into the contractive
  semigroup (generalized  diffusion equation) dynamics  (23) and (24),  by means of an analytic continuation in time.
  This mapping we  reproduce in the book-keeping (Wick rotation) form; $(it) \rightarrow t$.

  We emphasize that   one should  not mystify  the  emergent "imaginary time"  transformation.
  The time label $t$ pertains to  fairly   standard
  (unequivocally real)  dynamics scenarios. However the  detailed   patterns of temporal behavior
   for two   motion scenarios   are  very different.

  Nonetheless, various  (real) functionals of dynamical
  variables, evolving in accordance with the chosen  motion rule (say, Schr\"{o}dinger's $\psi (t)$ and$ \psi ^*(t)$,
    may  be consistently transformed  into the affiliated functionals
  that evolve according to another,  we call it \it  dual, \rm  (dissipative  $\theta ^*(t)$ and  $\theta (t)$) motion rule.
  The reverse transformation works as well.
  We shall address  this issue in below, from   varied perspectives.

\section{Scale covariant  patterns of evolution}

Let $[\Delta l]$ stand for an arbitrary (albeit fixed for the present purpose) unit of length.
For a continuous probability density  $\rho $ on $R^1$ we can introduce its   (dimensionless)
Shannon entropy functional, also named differential entropy:
\cite{gar2}
\begin{equation}
S(\rho ) \doteq   -   \int d\left( {\frac{x}{[\Delta l]}}\right) \, ([\Delta l]\rho )\, ln ([
\Delta l]\rho )   =  - \int   dx\,\rho \, \ln ([\Delta l]\rho )\, .
\end{equation}
If $\rho (x,t)$ depends on time, then $S (\rho )=S(t)$ may evolve in time as well.

The time rate equation for $S(t)$ is devoid of any  $[\Delta l]$ input and  for all $\kappa \geq 0$ has the very same
 functional  form:
\begin{equation}
D \dot{S}= D\{S, H_{\kappa } \} =    - \langle u\, v \rangle \label{rate}
\end{equation}
 Obviously, $\langle \cdot \rangle $ stands for   the mean value with respect to a probability density $\rho (x,t)$.

 The above time rate formula does not depend on a specific unit of length, that is present in the definition  of the
dimensionless Shannon entropy.
However, the  entropy functional  itself \it is \rm  sensitive to scaling transformations.

Setting $x'= x/\beta $,  where $\beta >0$ is the scale parameter,   we have
$1 =\int \rho (x) \, dx= \int  \beta  \, \rho  (\beta  x')\, dx' $. Accordingly, the scale transformation $x\rightarrow  x'= x/\beta $
induces a transformation of the probability density in question:
\begin{equation}
\rho (x) \rightarrow \beta \, \rho (\beta x')\doteq \rho '(x') \, .
\end{equation}

Given the  Shannon entropy of the density  $\rho (x)$, we can  compare the outcome with that for the density $\rho '(x')$:
\begin{equation}
S'(\rho ')  \doteq - \int dx' \, \rho '(x') \ln \left([\Delta l]\rho '(x')\right)
  =  - \int dx\, \rho (x) \ln \left(\beta \, [\Delta l] \rho (x)\right)  =
 S(\rho ) - \ln \beta  \, .
\end{equation}
Consequently, the $x\rightarrow x/\beta $  scaling is equivalent to the sole change of the length unit
 $[\Delta l]  \rightarrow   \beta \, [\Delta l]$  in the Shannon entropy definition.

Since $\beta $ is a fixed scaling parameter, the time rate formula (\ref{rate}) is scale independent.
Note that  by setting $\beta \doteq \exp \alpha $, we get $S'(\rho ')  = S(\rho ) - \alpha $.

We observe that  $u'(x') = \beta \, u(x)$ and the scale invariance of $D \dot{\cal{S}}= - \langle u\, v \rangle $ tells us that
$v'(x')= (1/\beta )v(x)$.  Since we assume $v(x,t)$ to be  the  gradient field, $v= (1/m)\nabla s $, we readily  arrive at the
corresponding
scaling property of $s(x,t)$:
\begin{equation}
s'(x') = {\frac{1}{\beta ^2}} s(x) \longrightarrow   v'(x') = {\frac{1}{\beta }}v(x) \, .
\end{equation}

We demand furthermore that the  ($\kappa =1$)  equations  $\partial _t\rho = - \nabla (\rho \, \cdot  v)$  and
$\partial _t s + (1/2m) (\nabla s)^2  +   V =0$  are scale invariant,  e.g. they retain their  form
after introducing $x'$, $\rho '(x',t)$ and $s'(x',t)$  instead $x$, $\rho (x,t)$ and $s(x,t)$ respectively.
This implies an  \it induced  \rm  scaling  property  of $V(x)\rightarrow V'(x') $:
\begin{equation}
V'(x') = {\frac{1}{\beta ^2}} V(x)
\end{equation}
and thus $H'_1= (1/\beta ^2) H_1$.

In view of
\begin{equation}
Q'(x',t)= \beta ^2 Q(x,t)
\end{equation}
the   general   evolution equations    for $\kappa \neq 1$   are \it not \rm scale invariant.
Let us consider a cumulative effect of the above scaling rules upon  the Hamiltonian
\begin{equation}
H_{\kappa }= \int dx \rho  [ (mv^2/2) + V + (1-\kappa )(mu^2/2)]
\end{equation}
according to:
\begin{equation}
H_{\kappa }\doteq H_{\kappa }(t)  \rightarrow H'_{\kappa }(t) =
 \int dx'\, {\cal{H}}(\rho' ,s')(x',t) \, .
 \end{equation}
Following the guess of
\cite{brenig}, originally  analyzed in  the  case of $V\equiv 0$, we take $\beta =\exp (\alpha /2) $; there
follows
\begin{equation}
H'_{\kappa }(\alpha ) =     \exp( -\alpha  )  \int  dx\, \rho \,  [{\frac{m}{2}}v^2 + V]  +
\exp(+ \alpha ) \int dx  \, \rho \, (1-\kappa )
{\frac{m}{2}}u^2
\end{equation}
which can be recast as the hyperbolic transformation with a hyperbolic angle $\alpha $
\begin{equation}
H'_{\kappa }(\alpha ) =  \cosh\alpha \, H_{\kappa }  - \sinh\alpha \, K_{\kappa } \,.  \label{first}
\end{equation}
Here, in addition to $H_{\alpha }$, we encounter a new  Hamiltonian  generator $K_{\kappa }$
 \begin{equation}
 K_{\kappa }  \doteq   \int dx\, \rho \left[{\frac{m}{2}}\,  v^2\,    +  V    - (1 -  \kappa )\, {\frac{m}2}\, u^2\right] \, ,
\end{equation}
The  difference  between  $H_{\kappa }$ and $K_{\kappa }$    is  encoded in the  sole  sign inversion of the last $(1-\kappa )(mu^2/2)$
entry.

We can  as well analyze an effect of the scaling transformation upon $K_{\kappa }$, \cite{brenig}:
\begin{equation}
K'_{\kappa }(\alpha ) = - \sinh \alpha  H_{\kappa } + \cosh \alpha K_{\kappa }\, . \label{second}
\end{equation}

Hyperbolic rotations that are explicit in equations (\ref{first}) and (\ref{second})  form a conspicuous
 Lorentz-type transformation  (hyperbolic rotation  of coordinates in Minkowski space) of
the direct  analogue  $(H_{\kappa }, K_{\kappa },0,0)$  of the familiar   Minkowski space  vectors  $(p_0,p,0,0)$ and/or   $(ct,x,0,0)$.

We recall that the time label $t$ is left untouched by hitherto considered scale transformations.
Consequently, like $H_{\kappa }$, the  generator $K_{\kappa }$ and  the induced  $\alpha $-family of
  generators $H'_{\kappa }(\alpha )$, $K'_{\kappa }(\alpha )$
  are legitimate  Hamiltonian  generators of \it  diverse \rm  time evolution scenarios, all running   with respect to
  the same  time  variable  $t$.
  The pertinent motions arise
  through  common  for all Hamiltonians,   a priori prescribed,  symplectic structure  Eq.~(\ref{bracket}) and the generic
  time evolution  rule (\ref{recipe}).

{\bf Remark 1:}  By re-tracing back the passage from the modular equation to the corresponding  Lagrangian and Hamiltonian, we realize that
$K_{\kappa }$  can be associated with  another $(\kappa -2 )$-family ($\kappa \geq 0$) of modular Schr\"{o}dinger equations:
\begin{equation}
i\hbar \partial _{\tau }\psi = [-(\hbar ^2/2m) \Delta + V]\psi  +(\kappa -2)Q\, \psi \, .
\end{equation}
to be compared with the originally introduced  $\kappa $-family
\begin{equation}
i\hbar \partial _t\psi = [-(\hbar ^2/2m) \Delta + V]\psi  - \kappa Q\, \psi \, .
\end{equation}
We shall not be elaborate on the general $\kappa > 2$ parameter regime   and, in view of the previously established reduction
 procedure (Section 1.3), we
confine our further discussion to  specific $\kappa =0$, $1$ or $2$ cases.

{\bf Remark 2:}  Let us notice that $K_0 \equiv  H_2$, $H_1 \equiv K_1$
 and $K_2 \equiv H_0$. We have   $H'_1=
\exp(-\alpha ) H'_1$  and the following   hyperbolic  transformation properties  hold true for  $H_0$ and  $H_2$:
\begin{equation}
H'_0 = \cosh \alpha H_0  - \sinh \alpha H_2
\end{equation}
and
\begin{equation}
H'_2 = - \sinh \alpha H_0  +  \cosh \alpha H_2\, .
\end{equation}
These transformation rules are a generalization of those presented for the $V \equiv 0$ case by L. Brenig, \cite{brenig}.

{\bf Remark 3:}  Hamiltonians of the form
\begin{equation}
H_0 = (m/2) \langle v^2 + u^2 \rangle + \langle V\rangle
\end{equation}
 and
\begin{equation}
 H_2= (m/2)\langle v^2-u^2 \rangle + \langle V\rangle
 \end{equation}
 are known in the literature, \cite{hasegawa,skorobogatov}, and  are interpreted to set  quantum-mechanical  and dissipative-dynamical
 frameworks respectively.
 Some elementary hints in this connection can also  be found in \cite{brenig,gar,gar1,gar2}.
Coming back to our observation that  a rescaling of the dimensional unit $[\Delta l]\rightarrow \exp \alpha [\Delta l]$  induces a transformation
$S'(\rho ')= S(\rho )- \alpha $ of  the Shannon entropy  $S(\rho )$   for a continuous probability density $\rho $, we realize that the choice of
$\alpha  \leq 0 $ implies a sharpening of the resolution  unit, hence  an effective growth of the Shannon entropy with the   lowering
of   $\alpha = -|\alpha  |$.
Indeed, we have  $S'(\rho ')= S(\rho )+ | \alpha |$. In this situation, effective (rescaled form) Hamiltonians  $H'_0$ and   $H'_2$,
  always have  an admixture of both
non-dissipative and dissipative components,  $H_0$ and $H_2$ respectively.

\section{Dynamical duality}

 Remembering that  for general functionals $G(t)$ of $\rho $ and $s$, we have   $\dot{G} = \{G,H_{\kappa }\}$, let us  consider a product
  ${\cal{F}} (x,t)  \doteq  - \rho (x,t)\, s(x,t)$ of  conjugate fields $s$ and  $\pi _s = - \rho $.
 The time evolution of
\begin{equation}
F(t)= \int dx\, {\cal{F}}(x,t) \doteq   -  \langle s\rangle
\end{equation}
looks quite interesting:
 \begin{equation}
 {\frac{d F}{dt}} =  \{ F,\, H_{\kappa}\} =  - \int dx\, \left[ s(x,t) \, {\frac{\delta H_{\kappa}}{\delta s}}  -
  \rho (x,t)\, {\frac{\delta H_{\kappa}}{\delta \rho }} \right]= \, \label{rule}
 \end{equation}
 $$
 - \int dx\, \rho \, \left[ {\frac{m}2} v^2 - V - (1-\kappa) {\frac{m}2} \, u^2\right] \, .
 $$
A new  Hamiltonian-type functional  has emerged on the right-hand-side of the dynamical identity (\ref{rule}).
 Let us denote
 \begin{equation}
H^{\pm }_{\kappa }  =      \int dx\, \rho \, \left[ {\frac{m}2} v^2 \pm  V   \pm  (1-\kappa) {\frac{m}2} \, u^2\right] \, .
\end{equation}
We  point out that in contrast to  previous scaling-induced Hamiltonians, the negative sign has been generated \it both \rm with respect to terms
$(m/2)\langle u^2 \rangle$  and $\langle V\rangle $.  The $+V \rightarrow  -V$ mapping was painfully lacking in the previous discussion to have
 properly implemented,  the  implicit,  analytic continuation in time procedure.

 The previous motion rule  rewrites as
 \begin{equation}
 {\frac{dF}{dt}} = \{ F, H^+_{\kappa } \} = -  H^-_{\kappa }(t)\,  ,
 \end{equation}
  where  $H^+_{\kappa }=H $ is the time evolution generator  Eq.~(\ref{hamiltonian}). Note that $H^+_{\kappa }$ is here a constant
   of motion, while $H^-_{\kappa }(t)$ is \it not. \rm

  After accounting for the Poisson bracket (\ref{bracket}),  we   encounter  a complementary relationship for  the
 time evolution that is generated by the  \it   induced \rm Hamiltonian      $H^-_{\kappa }$:
 \begin{equation}
  {\frac{dF}{dt }} = \{ F, H^-_{\kappa } \} = - H^+_{\kappa }(t)      \, .
 \end{equation}
Presently,    $H^+_{\kappa }$ is a constant  of motion, while    $H^-_{\kappa }(t)$ no longer is.

For \it each  \rm value of $\kappa \in R^+$, we  thus  arrive at  \it dual     \rm   time-evolution  scenarios,
  generated by  Hamiltonians $H^+_{\kappa }$ and  $H^-_{\kappa }$, respectively.  The \it duality \rm notion stays in conformity
   with our previous discussion  (Section 1.3)  of  the "imaginary time" transformation.

These  dual  motions, even  if started from the very same initial $t=0$  data  and allowed to continue indefinitely,
do   result in disparately different patterns of behavior: non-dissipative and dissipative, respectively.

Nonetheless, if we admit that the  a priori chosen generator of motion is $H^+_{\kappa }$, then we are interested
as well in:
\begin{equation}
{\frac{d^2 F}{dt^2}}=  -\{H^-_{\kappa }(t),H^+_{\kappa } \}  = -{\frac{dH^-_{\kappa }}{dt}}=  +2  \int   \rho \, v\,
\nabla [V+ (1-\kappa )Q] \, dx \label{standard}
\end{equation}
while, presuming that $H^-_{\kappa }$ actually  is the evolution generator, in the dual evolution  formula
\begin{equation}
{\frac{d^2 F}{{dt}^2}}= - \{H^+_{\kappa }(t),H^-_{\kappa } \}  = - {\frac{dH^+_{\kappa }}{d\tau }}= - 2
\int  \rho \, v\, \nabla [V+ (1-\kappa )Q] \,  dx \,  . \label{dual}
\end{equation}

The right-hand-sides of Eqs. (\ref{standard}), (\ref{dual}) associate  the    mean power   transfer rates (gain, loss or  none)  to
 the  dual  Hamiltonian evolutions of  $F(t)$,  generated by  $H^+_{\kappa }$   and  $ H^-_{\kappa } $ respectively.
 This   remains  divorced from the fact
 that  $H_{\kappa }^+(t)$ and $H_{\kappa }^-(\tau )$ actually   are constants of the pertinent dual  motions.

For clarity of  our  discussion we shall confine further  attention to $L^2(R^n)$ solutions of (1) and  (3). We  assume
that the external potential $V(x)$  is a continuous function that is bounded from below. If
 the energy operator  $\hat{H}= - (\hbar ^2/2m)\Delta + V$   is self-adjoint,
 then  we  have consistently   defined unitary transformations $\exp(-i\hat{H}t/\hbar )$  in $L^2(R)$ so that
  $i\hbar \partial _t \psi = \hat{H}\psi$
  holds true ($\kappa=0$).

The case of $\kappa=1$ we associate with two classes of external potentials $\pm V(x)$, with $+V(x)$ bounded from below.
This will  allow us to discriminate between the confining and scattering regimes. The "borderline" meaning of $\kappa=1$
can be read out from Eq.~(\ref{lagr}), where $mu^2/2$  contributions  cancel away.

In conformity with our previous discussion of generalized heat equations, related to solutions   of  Eq.~(\ref{one}) with $\kappa=2$,
 given  $+V(x)$  and $\hat{H}$,  we  pass to  a pair of time-adjoint  parabolic  equations:
  $\hbar \partial _{t} \theta _*  =  - \hat{H}  \theta _*$ and
  $\hbar \partial _{t} \theta  = \hat{H}  \theta $.

  Here,  $\theta _*(x,t) = [\exp(-\hat{H}t/\hbar )\, \theta _*](x,0)$ represents a  forward  dynamical semigroup evolution,
  while $\theta (x,T-t) = \exp(+\hat{H}t/\hbar ) \, \theta (x,T)$  stands for a backward one. Both are  unambiguously defined
  in a finite time interval $[0,T]$, provided one has prescribed  suitable  end-point data \cite{zambrini}.

  The corresponding   modular Schr\"{o}dinger equations (plus their complex conjugate versions)   read:\\
 (i)  $\kappa=0$ $\Longrightarrow $ $i\hbar \partial _t \psi = [-(\hbar ^2/2m)\Delta + V] \psi$\\
 (ii) $\kappa=1 $ $\Longrightarrow $ $i\hbar \partial _t \psi = [-(\hbar ^2/2m)\Delta  \pm  V   -   Q  ] \psi$\\
  (iii) $\kappa=2$ $\Longrightarrow $ $i\hbar \partial _t \psi = [-(\hbar ^2/2m)\Delta  -   V - 2 Q ] \psi$.

The  associated  Lagrangian densities (\ref{lagr})  in the $(\rho ,s)$-representation  and the induced dynamical rules
 are worth listing as well. In addition to the continuity equation  $\partial _t\rho = - \nabla (\rho \, \cdot  v)$ we have
 valid the Hamilton-Jacobi type equations:  \\
(i)   $\kappa=0$;    $   {\cal{L}}  =  - \rho \left[ \, \partial _ts
      + (m/2)(v^2 + u^2)  +  V\right] $ $\Longrightarrow $   $\partial _t s + (1/2m) (\nabla s)^2  + (V + Q)=0$ \\
(ii)  $\kappa=1$;   $   {\cal{L}}  =  - \rho \left[ \, \partial _ts
      + (m/2)v^2     \pm  V \right] $  $\Longrightarrow $ $\partial _t s + (1/2m) (\nabla s)^2  \pm V =0$ \\
(iii) $\kappa=2$;  $ {\cal{L}} =  - \rho \left[ \, \partial _ts
      + (m/2)(v^2  -  u^2) - V\right]  $ $\Longrightarrow $ $\partial _t s + (1/2m) (\nabla s)^2  - (V + Q)=0$.

      On dynamically admitted fields  $\rho (t)$ and $s(x,t)$,   $L(t) \equiv 0$,  i. e.
$\langle \partial _t s\rangle  =- H $.
The respective  Hamiltonians   (\ref{hamiltonian}) do follow:\\
(i)  $ H^+ \doteq \int  dx \, \rho \left[(m/2) v^2       +  V   + (m/2) u^2  \right] $\\
(ii) $H_{cl}^{\pm} \doteq  \int  dx \, \rho \left[(m/2) v^2    \pm  V  \right] $\\
(iii) $H^-  \doteq \int  dx \, \rho \left[(m/2) v^2   -  V   -(m/2) u^2  \right] $

We  emphasize   that, from  the start,
  $V(x)$  is chosen  to be a   continuous and bounded from below function.
 In the   definition of the above  Hamiltonians there is no $\kappa $ label anymore  and  a subscript "cl" refers to the classically
 motivated (Hamilton-Jacobi  theory) wave formalism.

  The   evolution equations for $F=-\langle s\rangle $, c.f. \cite{brenig} and \cite{gar1,gar2},  clearly define
   \it dual  \rm pairs:
\begin{equation}
\dot{F} =  \{ F,\, H^+\}  =- \int dx\, \rho \, \left[ {\frac{m}2} v^2 - V - {\frac{m}2} \, u^2\right]  =  -H^- (t)\, ,
\end{equation}
\begin{equation}
\dot{F} =  \{ F,\, H^-\} =  -\int dx\, \rho \, \left[ {\frac{m}2} v^2  +  V + {\frac{m}2} \, u^2\right]= -  H^+(t) \, \label{diff}
\end{equation}
and
\begin{equation}
\dot{F} =  \{ F,\, H_{cl}^+ \} = -\int dx\, \rho \, \left[ {\frac{m}2} v^2  -   V \right]  = - H_{cl}^-(t)
\end{equation}
\begin{equation}
\dot{F} =  \{ F,\, H_{cl}^- \} =  -\int dx\, \rho \, \left[ {\frac{m}2} v^2  +  V \right]  = - H_{cl}^+(t) \, . \label{diffcl}
\end{equation}
It is instructive to notice that the functional $F(t)$ in Eqs.~(\ref{diff}) and (\ref{diffcl}) may consistently play
 the role of a Lyapunov functional, indicating the preferred sense of time  ("time arrow")  in the course of the evolution process.
 Namely, if we take $\langle V\rangle >0$, the right-hand-side  expression is  negative definite. Hence $F(t)$ is a monotonically decaying
  function of time  which is a standard signature of  a dissipation process, c.f.  the Helmholtz free energy
  and  the  relative entropy discussion  for diffusion-type processes, \cite{gar1,gar2,gar3}.

In the notation of section 1, c.f.   Eq.~(\ref{hamiltonian}),  we have $H^{\pm } = H_0^{\pm }$   and $H^{\pm }_{cl} = H_1^{\pm }$.
The corresponding  (dual)  time  rate formulas (\ref{standard}), (\ref{dual}) do follow.

The motion rules for $\dot{F}(t)$ can be given   more transparent form by reintroducing  constants $H^{\pm }$ of the  respective motions.
Then
\begin{equation}
 \dot{F}(t) = -m \langle v^2\rangle (t)  +  H^{\pm }  \label{lyapunov}
  \end{equation}
 and
    \begin{equation}
 \dot{F}(t )= -m \langle v^2\rangle (t )  +  H^{\pm }_{cl}  \, .
\end{equation}
Here, the   non-negative  term   $m \langle v^2\rangle (t)$ should receive due attention, because  of its utmost importance in the study of
the Shannon entropy dynamics, \cite{gar1,gar2,gar3}, where it represents an \it entropy production \rm time rate. The latter
 is generated solely by the  dynamical processes  which are  intrinsic to the system
 and  does not involve  any  energy/heat flow,  in or out of the  potentially dissipative  system (that would need the  notion
 of an external to the system, thermal reservoir).

Since  $H^+$   and  $H^- $  are constants of respective motions,   $F(t) - t\, H^{\pm }$   are monotonically decreasing
in    time  quantities. This property extends to the $H^{\pm}_{cl}$  generated dynamics  as well.

   The speed (slowing down, or acceleration), with which the above  decay process  may occur, relies on  the specific  dynamical pattern of
   behavior  of
   $\langle v^2 \rangle $. That  is quantified by $- m d\langle v^2\rangle /dt$, hence:
\begin{equation}
 \ddot{F}(t)= \{\dot{F}(t), H^{\pm }\} =   \pm   2 \int \rho v \nabla (V+Q) dx \, .
\end{equation}
   For $H^{\pm }_{cl}$ generated motions, we have:
\begin{equation}
  \ddot{F}(t)= \{\dot{F}(t), H^{\pm }_{cl}\} = \pm     2 \int \rho v \nabla V dx  \, .
\end{equation}

We point out that a  major distinction between the dual  dynamical rules is encoded in the right-hand-sides of
   the above equations: the  conspicuous sign inversion is worth contemplation. The related integrals have a
    clear  meaning of the power transfer (release,  absorption or  possibly none), in the mean, that is induced
    by  time evolution of  the pertinent
    dynamical system, \cite{gar1,gar2}.

Let us stress that the   "imaginary time" transformation, even if  not  quite  explicit in our discussion,
 hereby has been  extended  to  dynamical  models of purely classical provenance. The pertinent Wick rotation
  connects  confined and scattering   motions, admitted to  occur  in  a continuous and  bounded from below  (confining)
 potential $+V(x)$ and  in its  inverted (scattering)  counterpart $-V(x)$, respectively.  An obvious example of an inverted
  harmonic oscillator \cite{barton} is worth mentioning at this point. Parabolic potential barriers \cite{shimbori}
   and repulsive $1/r^2$ potentials   \cite{moritz} belong to the same category.

 In the above  discussion, the  sign  inversion issue is manifested in   second  time  derivatives of various functionals.
In view of  $\pm  \nabla V$ presence,  we can identify this behavior as a remnant of  the standard classical (Newtonian)
 reasoning, \cite{zambrini}:  if the sign looks wrong with respect to the classical Newton equation (e. g.we have $+\nabla V$),
 we can correct this "defect" by interpreting time $t$ as an "imaginary time" $it$ (or $i\tau $ to avoid a notational confusion).

\section{Physics-related implementations of the dual  dynamics patterns: a (dis)illusion  of  an "imaginary time"}

\subsection{Generalities}

 With the notational conventions $D=\hbar /2m$,
$b(x,t)\doteq v(x,t) + u(x,t)$, while imposing   suitable boundary conditions (e.g. $\rho , b\rho , v\rho $ vanishing at integration
boundaries or infinities),  we can write the Shannon entropy time rate of change in a number of equivalent  ways, \cite{gar2}.
\begin{equation}
D \dot{S}= D\{S, H_{\kappa } \} =    - \langle u\, v \rangle = \langle v^2\rangle - \langle b\, v\rangle
\end{equation}
and
\begin{equation}
D \dot{S}  = D\langle \nabla v\rangle = \langle \nabla b \rangle + D\langle u^2\rangle \, .
\end{equation}
  The  non-negative entry  $(1/D) \langle v^2\rangle $ is interpreted
as the \it  entropy  production \rm rate in the considered dynamical  system \cite{gar1,gar2,gar3}.

We note that
\begin{equation}
D\langle u^2 \rangle = - D \langle \nabla u\rangle  = {\frac{2}m} \langle Q \rangle
\end{equation}
so that the mean divergence of an osmotic velocity is always negative. Here  $\langle Q\rangle  >0$ holds true for all finite times, \cite{gar2}.
 The  value $0$ can be achieved, if at all,  only in the asymptotic regime $t \rightarrow \infty $.

 To give a flavor of the time duality (and specifically the "imaginary time transformation") connection, let us mention that the free
  Brownian motion can be embedded in the   above  scheme by setting $b\equiv 0$ and regarding $D$ as the diffusion constant.
 We have   $v = - u = - D\nabla \ln \rho $  and  Shannon entropy time rate takes the form of  the
  de Bruijn identity $D\dot{S} =  \langle v^2\rangle = \langle u^2\rangle $, \cite{gar2}.

  The Shannon entropy $S(t)$ of the Brownian motion
    grows monotonically in time, solely due  to the entropy production $(1/D)\langle v^2\rangle (t)$.   The latter, in turn, is known to be a
    decreasing function of time (at least in the large time asymptotic) and ultimately is bound to vanish at  $t\rightarrow \infty $.

    In case of the diffusion process in a conservative potential  it is  not Shannon entropy, but  the Helmholtz free energy that  takes
    the role of the Lyapunov functional  and sets the "time arrow", c.f. Eqs.~(\ref{lyapunov} and    see   \cite{gar2,gar3}.
    Note that $F(t) - tH^-$ is a monotonically decaying in time quantity.

For comparison,  the Schr\"{o}dinger picture   quantum dynamics typically  involves   $b \neq 0$. In the special case of  the  free motion
 $\dot{S}>0$,  hence  $S(t)$ grows indefinitely,  \cite{gar1}.   In the large time asymptotic
$\dot{S}\rightarrow 0$,   while the entropy production   $(1/D)\langle v^2\rangle$  remains untamed and  never vanishes  while
 approaching  a finite  positive  value.

  However, the would-be natural property $\dot{S}(t)>0$ is not generic  for  the quantum motion  in external potentials, \cite{gar4}.
  Nonetheless $F(t)  - tH^+ $   does monotonically decrease with   time $t\rightarrow  \infty $ indicating  the Lyapunov
  functional-induced "arrow of time".  Even, though  this dynamics is manifestly non-dissipative.

\subsection{Harmonic oscillator and its inverted partner}

Let us consider a standard  classical  harmonic oscillator problem,  where
\begin{equation}
H \doteq {\frac{p^2}{2m}}  +  {\frac{1}2} m\omega ^2 q^2
\end{equation}
is an obvious constant of motion for the  Newtonian system $\dot{p}= m\ddot{q} =  - m\omega ^2q$,
\begin{equation}
q(t) = q_0 \cos \omega t + {\frac{p_0}{m\omega }}\sin \omega t
\end{equation}
$$
p(t) = p_0 \cos \omega t - m\omega q_0 \sin \omega t \, .
$$
Clearly $H=  p^2_0/2m + (m\omega ^2/2)q^2_0$  is a positive constant.

Let us  perform an analytic continuation in time, by considering the Wick rotation $t\rightarrow  -it$, paralleled by
the transformation of initial  momentum data $p_0\rightarrow -ip_0$. Once inserted to the above
harmonic oscillator expressions, we arrive at
\begin{equation}
H_{-ip_0}  = -  p^2_0/2m + (m\omega ^2/2)q^2_0  \doteq - \overline{H}
\end{equation}
  and
\begin{equation}
q_{-ip_0}(-it) \doteq  \overline{q}(t) =    q_0 \cosh \omega t - {\frac{p_0}{m\omega }} \sinh \omega t
\end{equation}
together with
\begin{equation}
 p_{-ip_0}(-it) \doteq  + i\overline{p}(t) =   -i  p_0 \cosh \omega t + i m\omega q_0 \sinh \omega t \, ,
\end{equation}
which simply  rewrites as
\begin{equation}
\overline{p}(t) =  -  p_0 \cosh \omega t +   m\omega q_0 \sinh \omega t \, .
\end{equation}

We observe that
\begin{equation}
\overline{q}(-t) =  q_0 \cosh \omega t + {\frac{p_0}{m\omega }} \sinh \omega t
\end{equation}
and
\begin{equation}
 -\overline{p}(-t) =  p_0 \cosh \omega t +  m\omega q_0 \sinh \omega t
 \end{equation}
 are the  familiar inverted oscillator solutions, generated by $\overline{H}$, \cite{barton}.

Indeed, equations of motion for $\overline{q}(t)$  and $\overline{p}(t)$ directly derive from the
 Hamiltonian $H_{-ip_0}=- \overline{H}$ with
\begin{equation}
\overline{H} =  {\frac{\overline{p}^2}{2m}}    - {\frac{1}2} m\omega ^2 \overline{q}^2
\end{equation}
They give rise to the    Newton equation
$\dot{\overline{p}}= m\ddot{\overline{q}} =  + m\omega ^2\overline{q}$. However, the dynamics generated by $\overline{H}$
is related to that generated by $-\overline{H}$ by the time reflection:
 the latter dynamics runs backwards, if the former runs forward.

{\bf Remark:} At the first glance, the harmonic oscillator example may be regarded as   a  unique  special case  (linear dynamical system) to which
our imaginary time transformation arguments may be applied.  Fortunately, we can give a number of nonlinear models  whose (Euclidean) inversion can be
consistently implemented, \cite{rajaraman}.  The Euclidean connection goes beyond the confining vs scattering potential idea of ours and extends to periodic
 potentials as well. Examples from the instanton physics: static localized  (kink) solutions of the $\phi ^4$ nonlinear field theory in one space dimension may be
 interpreted as Euclidean  time solutions of the double well
potential problem;  the sine-Gordon kink may be interpreted as a  Euclidean time solution of a plane  pendulum problem.

\subsection{Time duality  via analytic continuation in time}

The above procedure gives clear hints  on how to connect the \it dual \rm  classical wave theory evolutions,
associated with  the  previously discussed
Hamiltonians  $H^{\pm }_{cl}$.  We recall that in addition to the continuity equation, we infer the dual
Hamilton-Jacobi equations  $\partial _t s + (1/2m) (\nabla s)^2  \pm V=0$ and  that
there holds  $\partial _t \rho = - \nabla \cdot (\rho \, v)$  with  $v(x,t) = (1/m) \nabla s(x,t)$.

In the adopted notational convention, we define  the initial data  $s_0(x)=- \overline{s}_0(x)$ and
introduce an "imaginary time"  transformation :
\begin{equation}
\psi (x,t) =\rho ^{1/2} \exp (is/2mD)  \longrightarrow
\overline{\psi }(x,t) \doteq   \psi _{-is_0}(x,-it)=   \label{wick}
\end{equation}
$$
\rho _{-is_0}^{1/2}(x,-it)  \exp [is_{-is_0}(x,-it)/2mD] \doteq
\overline{\rho }^{1/2}(x,t)  \exp [- \overline{s}(x,t)/2mD]  \, .
$$
We note that $\lim_{t\downarrow 0} \,  is_{-is_0}(x,-it) =  i(-is_0)(x,0)=s_0(x)$.

An analogous procedure for an analytic continuation in time  has been worked out  in the  general  context  of the
 Euclidean quantum  mechanics in Ref. \cite{zambrini1}, however with no  mention of its extension to the classically inspired
 Hamilton-Jacobi equation.

 Let us denote $\overline{v}= (1/m) \nabla \overline{s}$. Accordingly,  the transformations implemented by  (\ref{wick}) replace
 $H^+_{cl} = \int  dx\, \rho [(m/2) v^2 + V]$  by   $- {\overline{H}}\, _{cl}^- = \int  dx\,
 \overline{\rho }[- (m/2) \overline{v}^2 + V]$, with the very same function $V(x)$ in both expressions.

Clearly:
\begin{equation}
\partial _t \rho = - \nabla \cdot (\rho v)  \longrightarrow  \partial _t \overline{\rho } =
+\nabla \cdot (\overline{\rho }\,  \overline{v})
\end{equation}
which is an obvious indicative of the  time reflected (backwards) evolution.
Analogously
\begin{equation}
\partial _t s + (1/2m) (\nabla s)^2  \pm V=0  \longrightarrow  \partial _t\overline{s} - (1/2m) (\nabla \overline{s})^2 +V=0
\end{equation}
where, the time reflection $t\rightarrow -t$  induces an expected form of the dual   Hamilton-Jacobi equation:
\begin{equation}
\partial _t\overline{s} + (1/2m) (\nabla \overline{s})^2 -  V=0\, .
\end{equation}

The above discussed analytic continuation in time directly extends to the general pair $H^{\pm}$ of dual Hamiltonians, c.f.
Section  IV  of Ref. \cite{zambrini1}. The description becomes even more straightforward, because in this case we connect pairs of
linear partial differential equations.
If $\psi (x,t)$ actually is a solution of the Schr"{o}dinger  equation $i(2mD)\partial _t  =  \hat{H}\psi $, then
\begin{equation}
\psi _{-is_0}(x,-it)=\overline{\rho }^{1/2}(x,t)  \exp [- \overline{s}(x,t)/2mD]  \doteq \theta _*(x,t)
\end{equation}
solves a backwards diffusion-type equation
\begin{equation}
- (2mD) \partial _t \theta _* = \hat{H} \theta _*
\end{equation}
while
\begin{equation}
\theta (x,t) =\overline{\rho }^{1/2}(x,t)  \exp [+\overline{s}(x,t)/2mD]
\end{equation}
solves the forward equation
\begin{equation}
(2mD) \partial _t \theta  = \hat{H} \theta  \, .
\end{equation}
In the above one may obviously identify $D=\hbar /2m$.

The whole procedure can inverted and we can trace back a non-dissipative  quantum dynamics pattern which stays in affinity  (duality)
 with  a given dissipative dynamics, c.f. \cite{zambrini1}.

\subsection{Diffusion-type processes}

\subsubsection{Smoluchowski process}

The Hamiltonian appropriate for the description of dissipative processes (strictly speaking, diffusion-type stochastic processes)
 has the form
\begin{equation}
H^-  \doteq \int  dx \, \rho \left[(m/2) v^2   -  V   -(m/2) u^2  \right]
\end{equation}
with the a priori chosen, continuous and bounded from below potential $V(x)$.
It is the functional form of $V(x)$ which determines local characteristics of the  diffusion process, \cite{gar2}.

Once the Fokker-Planck  equation is inferred
\begin{equation}
\partial _t\rho =  D \Delta \rho - \nabla (b\cdot  \rho )  \, ,
\end{equation}
where  $\rho _0(x)$ stands for the initial condition,  we  adopt  the forward drift  $b=f/m\gamma $ of the process in  the  standard
Smoluchowski form, characteristic for  the Brownian motion in an external force field $f(x)= - \nabla {\cal{V}}$. Here, $\gamma $ is
a friction (damping) parameter  and, instead of $D=\hbar /2m$,  we  prefer to think in terms of
  $D= k_BT/m\gamma $ where $T$ stands for an  (equilibrium)  temperature of the reservoir.

An  admissible  form  of   ${\cal{V}} \rightarrow f= -\nabla {\cal{V}}$   must  be compatible with the Riccatti-type equation, provided  the
potential function  $V(x)$  has been a priori chosen:
\begin{equation}
V(x) =  m\left[ {\frac{1}2} \left(  {\frac{f}{m\gamma }}\right)^2  +  D\nabla \cdot \left(  {\frac{f}{m\gamma }}\right)\right] \, .
\end{equation}

The Fokker-Planck equation can rewritten as a continuity  equation  $\partial _t\rho = - \nabla \cdot j $  with the  diffusion current $j$ in the form:
\begin{equation}
j = \rho v = {\frac{\rho }{m\gamma }} [ f - k_BT \nabla \ln \rho ] \doteq   {\frac{\rho }{m}}\nabla s  \, .
\end{equation}
We recall  the  general  definition of the current velocity $v= (1/m) \nabla s$.

 Since the time-independent $s=s(x)$ is here admissible, we  have actually  determined
 \begin{equation}
s  =  -  {\frac{1}{\gamma }} ( {\cal{V}} +   k_BT \ln \rho )
\end{equation}
whose  negative   mean value  $F=-  \langle s\rangle $  determines  for   the    Helmholtz free  energy of
 the random  motion, as follows:
\begin{equation}
\Psi \doteq \gamma \,  F = U - T {\cal{S}} \, ,
\end{equation}
where  ${\cal{S}} \doteq k_B\,  S$   stands for  the  Gibbs-Shannon entropy of the continuous probability distribution, while  an internal energy reads
 $ U = \left< {\cal{V}} \right>$.

 Since we assume  $\rho $  and $\rho V v$ to vanish
at the integration volume boundaries,  we get
\begin{equation}
\dot{\Psi }  =     - (m\gamma )
 \left<{v}^2\right> = - k_BT (\dot{\cal{S}})_{int} \leq 0 \, . \label{helm}
\end{equation}
Clearly,  the Helmholtz free energy  $\Psi $ decreases as a function of time,
or  remains constant.

The  Shannon  entropy  $S(t)  = -\langle \ln \rho \rangle $
  typically is not a conserved quantity. We impose  boundary restrictions
that $\rho, v\rho, b\rho $ vanish  at spatial infinities or  other  integration interval borders  and consider:
\begin{equation}
  D \dot{S}  =  \left< {v}^2\right>
    -  \left\langle {b}\cdot {v}
 \right\rangle  \label{balance}  \, .
\end{equation}
which rewrites as follows
\begin{equation}
\dot{S} = (\dot{S})_{int} + (\dot{S})_{ext}
\end{equation}
where
\begin{equation}
k_BT  (\dot{S})_{int}  \doteq m\gamma \left<{v}^2\right> \geq 0
\end{equation}
stands for the {  entropy production} rate, while
\begin{equation}
k_BT  (\dot{S})_{ext}  =  -  \int {f} \cdot {j}\, dx = -
m\gamma \left\langle {b}\cdot {v}
 \right\rangle
\end{equation}
 (as long as negative  which is not a must)  may be  interpreted as the  {  heat dissipation rate}:$ - \int {f}\cdot {j}\,  dx$.

Let us consider the stationary regime   $\dot{S} =0$
associated with an (a priori assumed to exist)
invariant density $\rho _{*}$.
 Then,    $$b=u = D \nabla  \ln \rho _{*} $$ and
\begin{equation}
 -(1/k_BT)\nabla {\cal{V}} = \nabla \ln\, \rho _{*}  \Longrightarrow \rho _{*} = {\frac{1}Z} \exp[ - {\cal{V}}/k_BT]\, .
 \end{equation}
Hence
\begin{equation}
- \gamma s_*  = {\cal{V}} + k_BT \ln \rho _{*}  \Longrightarrow  \Psi _{*}  =
 - k_BT \ln Z  \doteq \gamma  F_{*}
 \end{equation}
   with   $Z= \int \exp(-{\cal{V}}/k_BT) dx$.
 $\Psi _*$ stands for   a  minimum  of  the time-dependent  Helmholtz
free  energy $\Psi $. Because of
\begin{equation}
Z= \exp (-\Psi _*/k_BT)
\end{equation}
 we have
\begin{equation}
\rho _* =
\exp[(\Psi _* - V)/k_BT] \, .
\end{equation}

Therefore,   the {  conditional  Kullback-Leibler   entropy} ${\cal{H}}_c$,
of the density $\rho $    relative to an equilibrium (stationary)  density $\rho _*
$ acquires the form
\begin{equation}
 k_BT {\cal{H}}_c  \doteq  - k_BT \int \rho \ln
({\frac{\rho }{\rho _*}})dx = \Psi _* - \Psi  \, .
\end{equation}

In view of the concavity property of
the function $f(w) = - w\ln w$,  ${\cal{H}}_c$ takes only
negative values, with  a maximum at $0$. We have   $\Psi _*\leq \Psi $ and
 $k_BT \dot{\cal{H}}_c = - \dot{\Psi }  \geq 0$.  ${\cal{H}}_c(t)$  is bound to  grow monotonically
 towards $0$,  while  $\Psi (t)$ drops down to   its
 minimum  $\Psi _*$  which is reached  upon  $\rho _*$.

 The Helmholtz free
 energy minimum   remains divorced from any extremal property of the Gibbs-Shannon
 entropy. Only the Kullback-Leibler  entropy shows up an expected  (growth)  asymptotic  behavior. See e.g. also \cite{gar2}.

Note that properties of the free Brownian motion can be easily inferred by setting $b\equiv 0$ in the above discussion.
Then, the diffusive  dynamics is sweeping and
there is no asymptotic invariant density, nor a finite minimum for $\Psi (t)$  which decreases indefinitely.

\subsubsection{Reintroducing duality}

To set a connection with the previous time duality ("imaginary time" transformation) framework, we need only to observe some
classic properties of Smoluchowski diffusion processes.
Once we set  $b= - 2D\nabla \Phi $  with $\Phi = \Phi (x)$,  a substitution:
\begin{equation}
\rho (x,t) \doteq \theta _*(x,t) \exp [- \Phi (x)]
\end{equation}
with $\theta _*$ and $\Phi $ being real functions, converts the Fokker-Planck equation
 into a generalized diffusion  equation for $\theta _*$:
 \begin{equation}
 \partial _t \theta _* = D \Delta \theta _* -  {\frac{V(x)}{2mD}}\theta _*
 \end{equation}
 and its  (here trivialized in view of the time-independence of $\Phi $) time adjoint
\begin{equation}
\partial _t \theta = -D\Delta \theta  + {\frac{V(x)}{2mD}}\theta
\end{equation}
  for a real function $\theta (x,t) = \exp [- \Phi (x)]$,
where
\begin{equation}
{\frac{V(x)}{2mD}} =  {\frac{1}2} ({\frac{b^2}{2D}} + \nabla \cdot b) =  D[(\nabla \Phi )^2 -\Delta \Phi ] \, . \label{fokkerpot}
\end{equation}
Let us note an obvious factorization property for the Fokker-Planck probability density:
\begin{equation}
\rho (x,t) = \theta (x,t)  \cdot \theta _*(x,t)
\end{equation}
which stays in affinity with  a quantum mechanical factorization formula $\rho =  \psi ^*  \psi $,  albeit presently
realized in terms of two  real functions $\theta $ and $\theta ^*$, instead of a complex conjugate pair.
In view of (at this  point we restore an original notation of Section  4):
\begin{equation}
\overline{\rho }^{1/2}(x,t) \,  \exp [- \overline{s}(x,t)/2mD]  \doteq \theta _*(x,t)
\end{equation}
we immediately recover
\begin{equation}
\overline{s} = (2mD)[ \Phi - (1/2) \ln \overline{\rho  }]
\end{equation}
in conformity with the previous definition Eq.~(83).  If there are no external forces, $\Phi $ disappears  and we are left with
the free Brownian motion associated  with  $\overline{s} = - mD\ln \overline{\rho }$.

For the record it is useful to mention explicit transformations between Green's functions appropriate for quantum motion and
transition probability densities  of standard diffusion type processes, \cite{wang}. Explicit examples of the free dual
 dynamics and those in the harmonic potential, have been worked out  there.\\

 \section{Conclusions}

  An analytic continuation in time (or an "imaginary time" transformation) stands for
   a mapping between two different  types of dynamics, both  running with respect to  the  equally "real"(istic) time.
   In our case, the modular Schr\"{o}dinger  equation has been a unifying  departure  point for an analysis of, sometimes not quite expected,
    affinities between  dynamical patterns of behavior  generated by the same  primary  nonlinear field, but typically considered
    disjointly - as research  problems  on  their own, for properly selected coupling parameter values.

    One may possibly take the view that   $\kappa =0, 1, 2$ cases are just  formally analogous   mathematical descriptions of  different
    physical systems. Our standpoint is that a primary \it  dynamical system  \rm  is a modular Schr\"{o}dinger equation with an  arbitrarily
    adjustable coupling parameter. Different coupling regimes refer to  \it   physically different \rm  patterns of behavior, but  there is a deep
    intertwine between  them, to be further explored.  The global changes of properties of solutions of nonlinear dynamical equations as the control
     parameter is varied are a routine wisdom for nonlinear system experts. This property is shared by the modular Schr\"{o}dinger equation as well.

     If one takes seriously   the dynamical duality (or  time duality) concept, the models
    considered  here should not be viewed independently  anymore. One can trace the
   dynamical patterns of one model in terms of those for another,
   and in reverse.  Even, if at the first glance,  the pertinent
   dynamics patterns may seem to have nothing in common.

    It is the Hamiltonian analysis of the nonlinear field (e.q. Schr\"{o}dinger equation) which has  led to  a variety of
    emergent  Hamiltonian motion scenarios.  Two classes of them, we
    call them confining or scattering, stay in  close  affinity on quantum, classical and stochastic (dissipative) dynamics levels of description.

   Euclidean methods are often used in  various  (especially quantum)  branches of the statistical physics research of equilibrium and
   near-equilibrium phenomena. Our discussion was
   basically concentrated on the real   time flow notion which, even after a Euclidean (imaginary time) transformation, still remains a real(istic) time
   flow, albeit  with a new physical meaning.

  It is worth mentioning that an an independent, quantum theory
  motivated approach  \cite{kazinski} (deformation quantization with an "imaginary
  transformed" deformation constant $\hbar $) has obvious links with
 the   Euclidean map  viewpoint towards diffusion-type processes,  explored in the present
 paper.

We would like to point out that, quite aside of existing and prospective  physical implementations,  the  sign-inversion issue for the
 conservative potentials has a deeper mathematical meaning  whose role we have slightly diminished, not to overburden the text with a strong
  admixture of an  advanced functional analysis. The dynamical semigroup indications in the present paper   were intended to   tell  the
 mathematically interested  reader under what circumstances one  can be sure of the existence (modulo
 suitable time interval limitations) of solutions that are connected by an imaginary time transformation.
 The semigroup notion cannot be hastily extended to the classical  dynamics, nonetheless an imaginary time link  still   works there.

{\bf Acknowledgement:}  I am indebted to Leon Brenig for pointing out to my attention his paper \cite{brenig} and enlightening
correspondence.\\

\end{document}